# Strain-induced transformation of charge-density waves and mechanical anomalies in the quasi one-dimensional conductors $TaS_3$ and $K_{0.3}MoO_3$.

S.G. Zybtsev, V.Ya. Pokrovskii[*]

*Kotel'nikov Institute of Radioengineering and Electronics of RAS, 125009 Mokhovaya 11-7, Moscow, Russia*

We report studies of low-field conductivity, $\sigma$, of the orthorhombic $TaS_3$ samples as a function of strain, $\varepsilon$. In the Peierls state the $\sigma(\varepsilon)$ dependencies show hysteresis. A similar hysteresis loop is observed for $K_{0.3}MoO_3$. For nano-sized $TaS_3$ samples the $\sigma(\varepsilon)$ curves show step-like changes associated with the "quantization" of the wave vector, $q$, of the charge-density wave (CDW). The dependences clearly reveal the change of the $q$-vector with strain. In contrast with the traditional concept, $q$ is found to *increase* with sample expansion. This means that the stretch-induced anomalies cannot be explained by the transition of the CDW to fourfold commensurability with the pristine lattice (lock-in transition). Alternatively, we suppose, that at the critical stretch a CDW with larger amplitude and modified $q$-vector forms. Further, the models describing metastable length states and drop of the Young modulus on CDW depinning in terms of longitudinal CDW strain, require reconsideration. Presumably, transverse effects should be taken into account.

## 1. INRTODUCTION

The quasi one-dimensional conductors, which are well-known firstly owing to the non-linear charge-density wave (CDW) transport [1], show also unique properties, which, to wide extent, can be called mechanical, namely: strain-induced features in conduction and thermopower [2,3,4,5,6], anomalous elastic [7,8,9,10,11,12,13,14], thermal-expansive [15] and electromechanical properties [16,17,18,19,20]. In particular, we can mention the abrupt changes in non-linear and linear transport under uniaxial stretch [3,4,5,6] accompanied with a feature in the stress-strain relation [14], drop of the Young modulus [7,8,9,11,12] and shear modulus [9,10,12,13] on the CDW depinning, hysteresis in thermal expansion [15], large electric-field-induced deformations (uniform [16] and non-uniform [17,18,19]). Generally speaking, all these features demonstrate interplay of the CDW and pristine-lattice properties.

The effects mentioned above are found in a number of compounds and clearly reveal general features of the CDW. At the same time, they reflect individual properties of each CDW compound. For example, several-percent drop of Young modulus on the CDW depinning has been reported for $TaS_3$ and $(TaSe_4)_2I$ [1,2 (p. 155)], a 1-2 orders of magnitude lower – for $NbSe_3$ [21,22], and no drop (to the accuracy of $5\times10^{-5}$) – for $K_{0.3}MoO_3$. [23]. Such a scattering of properties could be expected from general consideration (see the discussion in [13] in the end of p. 2970). Suppose, we take an ideal one-dimensional (1D) conductor, for which the CDW wave vector, $q$, is exactly equal to $2k_F$ ($k_F$ is the Fermi wave vector). Under stretch the linear concentration of electrons (per unit length of a conducting chain), $n_0$, falls as $1/c$, where $c$ is the lattice constant. This means that the equilibrium CDW wavelength $\lambda=\pi/k_F=2/n_0$ grows proportionally to $c$: $d\log \lambda/d\log c=1$. Thus, expansion of the sample does not lead to the CDW strain, and there is no ground for any anomalies. Therefore, the mechanical anomalies must originate from individual features of the compounds, such as charge transfer under strain [24,25,26], 3D effects, electron-hole asymmetry [27,28]. These effects can change $n_0$ under strain or make $\lambda$ deviate from $2/n_0$ [26,29]. In terms of the CDW wave vector $q$, this means that $qc \neq$ const.

---
[*] pok@cplire.ru

The non-trivial dependence of $q$ on strain, $\varepsilon$, gives the most common interpretation of the anomalies, at least of those concerning longitudinal strain. In this case the CDW-lattice coupling can be characterized in terms of the coefficient $g \equiv (d\log \lambda/d\log c) - 1 \equiv -d\log q/d\varepsilon - 1$ introduced in [29]. If $g=0$, no anomalies are expected.

The most detailed mechanics-related studies have been performed for TaS$_3$ [3-11,13-21], a typical Peierls quasi one-dimensional conductor with CDW forming at $T_P$=220 K [1]. For this compound an integral picture of the lattice-CDW interplay has been arranged. The basic point of this picture is that under uniaxial stretch the $q$-vector (namely, its longitudinal component) reduces; at a critical stretch, $\varepsilon_c$, $q$ achieves 4-fold commensurability with the lattice, i.e., shows a lock-in transition; at $\varepsilon > \varepsilon_c$ it is supposed to reduce further [3,4,5,6]. This inference has been based on a number of experiments: pronounced features in conductivity (both linear and non-linear) and thermopower are observed around $\varepsilon = \varepsilon_c$. The $q(T)$ dependence [1,30] seemed to confirm this picture: below $T_P$ $q$ is slightly above the 4-fold commensurability, but approaches it with $T$ decrease; the $\varepsilon_c(T)$ and $q(T)$ dependences were found to be qualitatively similar [3,4,15]. Confrontation of the dependences [15] gave $g=6$. The value and sign of $g$ was found to be consistent with the temperature hysteresis of length, $L(T)$ [15]. We will consider the connection of the loop $L(T)$ with $q(\varepsilon)$ dependence in the Discussion section of the Paper in detail.

The non-zero value of $g$ is also a key for explaining the softening of the lattice on the CDW depinning [13,26,29] at the electric field $E=E_t$. To the first approximation, the total elastic energy can be presented as a sum of the deformation energies of the lattice in itself and the CDW. According to the model [29], if $g \neq 0$, a longitudinal deformation of a crystal drives the CDW from the equilibrium (this is the condition of phase slippage absence). In the sliding state the deformation of the CDW relaxes through phase slippage, and the CDW elastic contribution drops out from the total energy. Thus Young modulus of the sample decreases.

As we mentioned above, no drop of the Young modulus was detected for K$_{0.3}$MoO$_3$ [23]. This "exception" seemed to confirm the overall picture: it was suggested that the bond in this compound is ionic, in contrast to trichalcogenides [24,25]. Thus, deformations do not change $n_0$, so $q/b^*$ [31] should be independent of $\varepsilon$ [13], and the CDW should not contribute to the total elastic energy.

Up to now, the self-consistent picture of mechanical anomalies in TaS$_3$ lacked one important element: direct evidence of the $q$ change under strain. The studies, presented in this paper, reveal the $q(\varepsilon)$ dependence. Contrary to the expectations, $q$ is found to ***increase*** with $\varepsilon$. A similar dependence is found for K$_{0.3}$MoO$_3$. These results are in contradiction with the picture described above and give ground for a serious reconsideration of the mechanical anomalies in the CDW compounds.

## II. EXPERIMENTAL APPROACHES

Structural studies under strain, including those at low temperatures, require an unconventional setting of the experiment and can be rather complicated. However, studies of the transport properties of the CDW conductors can give direct and precise information about the $q$ change. Two approaches are known, and both have been successful in recovering $q$ change with $T$, obtained from the diffraction experiments.

The first approach is based on the analyses of the hysteresis in the temperature dependence of conductivity, $\sigma(T)$. [27,28]. The conductivity is presented as a function of two parameters, $q$ and $T$. It is implied that $q$ can take different values (in a certain range around the equilibrium) at given $T$.

Below we are considering the changes of the CDW wave vector with respect to the lattice, therefore it will be reasonable to measure $q$ in $c^*$ units, i.e., introduce the dimensionless value $q^* \equiv q/c^*$. This normalization will be especially important below, when $q(\varepsilon)$ dependence will be considered.

For $TaS_3$ in a certain temperature range (approximately 100 K $<T<$ 140 K) the slope of the $\sigma(T)$ curve just after a reversal of temperature sweep is found to be much smaller than before the reversal. Clearly, this corresponds to the case when $q^*$ is (nearly) constant, because (nearly) no phase slippage occurs. In the mathematical form:

$$d\sigma/dT \ (\equiv \partial\sigma/\partial T|_{q^*} + \partial\sigma/\partial q^*|_T \, dq^*/dT) \gg \partial\sigma/\partial T|_{q^*}. \tag{1}$$

This means that the equilibrium $q$ **change** is coupled with $\sigma$ change as

$$d\sigma/dT \approx \partial\sigma/\partial q^*|_T \, dq^*/dT. \tag{2}$$

The simplest way to evaluate $\partial\sigma/\partial q^*|_T$ is to take the quasiparticle conductivity as unipolar: $p \gg n$, where $p$ and $n$ are hole and electron concentrations (per unit length of a conducting chain). This is a good approximation, at least, above ~90 K [32].

A change of the $q$-vector at given $T$ is connected with the difference of hole and electron excitations, $p$-$n$, as $\delta(p-n)/c^* = (1/\pi)\delta q^*$ [28]. Therefore, with $\sigma = ep\mu/s_0$, neglecting $n$ one obtains:

$$\partial\sigma/\partial q^*|_T = e\mu c^*/(\pi s_0), \tag{3}$$

or

$$\partial\log(\sigma)/\partial\log(q^*) \approx (\mu/\mu_{300})(\sigma_{300}/\sigma), \tag{3'}$$

where $\mu$ is the known [32] mobility of the hole-type quasiparticles (considered as independent of $q$), $e$ – the electron charge, $s_0$ – the area per one conducting chain (22 Å$^2$, see p. 399 in [1]). The relation (3') appears convenient in estimating $\partial\sigma/\partial q^*|_T$ from the experimental $\sigma(T)$ curve. Thus, with known $\sigma(T)$ one obtains $q^*(T)$. This consideration gives the dependence $[q^*(T)-q^*(0)]/q^*(0) \approx p(T)/n(300)$. This ratio can be estimated as $\sigma(300)/\sigma$, or, more exactly, from the ratio of Hall conductivities [27,28,32]. The resulting $q(T)$ dependence appears in quantitative agreement with the diffraction studies with the only fitting parameter $q(0)$. With the large scattering of the electron diffraction data for $TaS_3$ [30] the definition "quantitative" looks not very meaningful. However, similar consideration of $\sigma(T)$ hysteresis [28] in $K_{0.3}MoO_3$ gives $q(T)$ also consistent with the diffraction studies, which have been performed with X-ray technique and show much higher accuracy [33].

The second approach can be applied only to nano-sized samples, but can provide the $q^*$ change with extremely high accuracy and is independent of the compound parameters. It is known that nano-sized CDW samples show "quantization" of the $q$-vector [34,35,36], as well as the spin-density wave (SDW) compounds [37]. Discrete states correspond to different number of the wavelengths in a sample; transitions between the states (via phase slippage) occur as production or annihilation of a CDW (SDW) period, and can be detected directly with X-ray [37] or as step-like changes of conductivity [34,35,36]. In the second case the samples must be thin enough, so that the process would cover the whole cross-section, and short enough, so that the resulting $q^*$ change, $|\delta q^*|=c/L$ (or, $|\delta q^*|/q^*=\lambda/L$), would result in a detectable step of conductivity. Also, tight-boundary conditions for the CDW phase are required to be provided at the contacts. If these conditions are fulfilled, counting the number of steps in σ(T) one can find the $q^*$ change in the corresponding temperature range [38]. This approach provides an extremely high resolution in $q^*$ change, which can exceed that of the X-ray studies, like in the case of $K_{0.3}MoO_3$ [33,35].

Here both methods are applied for studies of the $q^*$ change in $TaS_3$. The key modification of the experiment is that instead of σ(T) we study σ(ε) curves. With this purpose we elaborated set-ups allowing closely continuous change of the sample length with high accuracy [39]. An additional requirement to the technique was that it should be applicable to nanosized samples, down to at least $s=10^{-3}$ μm² in cross-section and to $L=10$ μm. Therefore, we used the set-ups, in which the uniaxial stretch was achieved by means of bending a substrate. The sample is attached to one surface of the substrate made from an organic epoxy. The substrate is lying on two bearings at the ends, while the bending is provided by a bar (Fig. 1) driven from outside the cryostat by means of a mechanical motion transducer. The resulting strain is:

$$\varepsilon = 4\delta y d / L_{sub}^2, \qquad (4)$$

where $\delta y$ is the displacement of the bar, $d$ – thickness and $L_{sub}$ – the length of the substrate. A thin gold film playing the role of a strain gauge was deposited on the substrate near the sample. The strain-resistance coefficient (gauge factor) of the film was calibrated basing on the relation (4). The resolution of the strain control was well below $10^{-4}$. The maximum value of ε was about 1-1.5% and was limited by the substrate cracking.

…..

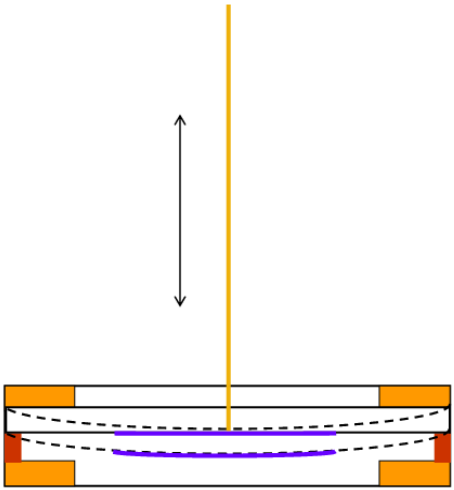

Fig. 1. A sketch of the substrate fixed over bearings (at the edges). Bending deformation is created by a bar. The bent substrate is drawn with broken lines. The sample is placed on the bottom surface of the substrate

To get temperature dependences of σ at approximately constant ε we used also the more common "lever" technique for stretching the samples [39]. The sample cross-section area was controlled with the help of RF interference (Shapiro steps) [40], basing on the ratio of the CDW current density at the 1st step to the irradiation frequency: $j_c/f=69$ A/MHz cm² for $TaS_3$. This control was especially important for short samples strained with the "lever" method. In this case the sample was partly covered with a shunting gold film, while only a short segment remained gold free. Below $T_P$ the film resistance was negligible. However, at room temperature we could not establish the sample cross section from the resistance. After the determination of the sample cross-section we could also subtract the contact resistance (see Fig 5 below), which can be relatively large at high temperatures Also, the control was important for the thinnest samples, whose resistivity can be larger [41].

## III. RESULTS.

Fig. 2 shows typical strain dependence of conductivity of a TaS$_3$ sample at $T$=123 K for relatively low values of ε. In agreement with [3,4,5] the conductivity grows with ε; the average slope of the curves ("gauge factor" of TaS$_3$), $d\ln\sigma/d\varepsilon \sim 100$, also agrees with the previous studies. The obvious feature of the dependence is the hysteresis. Similar hysteresis loops were observed previously in [10,42], though were not discussed. Application of a pulse $E > E_t$ changes the value of σ towards the center of the hysteresis loop (see the vertical arrows in Fig. 2). Thus, electric-field induces relaxation of the metastable states, like in the case of thermally induced metastability [1].

For larger strains, $\varepsilon = \varepsilon_c \sim 0.7$-$0.8\%$, σ achieves a maximum [3,4,5,39].

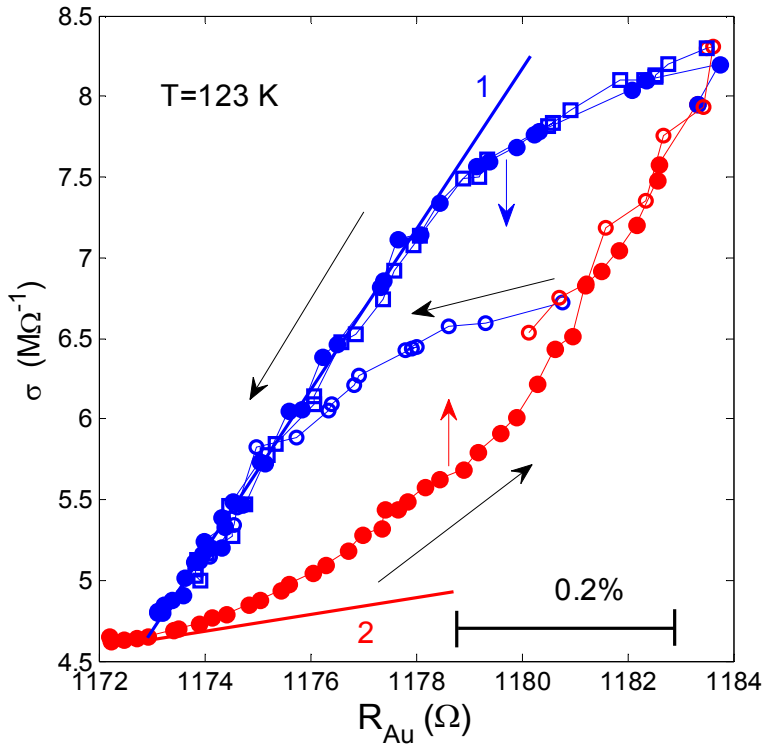

Fig. 2. Repeatedly recorded dependence of conductivity for a TaS$_3$ sample on strain (resistance of the gold film). Different markers correspond to different sweeps. The arrows indicate the direction of σ change after application of $E > E_t$. $L$=250 μm, $s$ = 0.4 μm$^2$.

The following analysis of the loop is closely similar with that of σ($T$) performed in [28] and Ref. 5 from [28]. One can notice that after reversal of stretching from decrease to increase (or vice versa) the slope of the σ(ε) curve reduces several times (compare the slopes of the straight lines 1 and 2). Like in the case of σ($T$) dependence, we attribute the hysteresis to $q$ falling behind its equilibrium value with ε change. Absence of phase slippage after the strain reversal is equivalent to $q^*$= const. Treating the σ(ε) loop similarly with the σ($T$) loop (Eqs. 1,2) [28] we come to

$$d\log(\sigma)/d\varepsilon \approx \partial\log(\sigma)/\partial\log(q^*)|_\varepsilon \, d\log(q^*)/d\varepsilon. \tag{5}$$

Thus the change of σ with ε (more exactly, 60-80% of it) is coupled basically with the change of $q^*$. One can see that $\partial\sigma/\partial q^*|_\varepsilon \, dq^*/d\varepsilon>0$. It is known that $\partial\sigma/\partial q^*|_\varepsilon >0$ [27,28,34]: increase of $q^*$ results in appearance of new free states under the Peierls gap, 2Δ, and thus, to increase of $p$. The sign of $\partial\sigma/\partial q^*|_\varepsilon$ can be easily checked also from the relaxation of thermally induced metastable states: application of $E>E_t$ after heating a sample results in an increase of σ (the red arrow in Fig. 2). Heating is known to result in increase of $q^*$ [1,30]. Before the relaxation $q^*$ was behind its equilibrium value, therefore, relaxation means increase of $q$, and this is what we intended to check.

With this, the first qualitative conclusion is that $dq^*/d\varepsilon>0$, i.e. *q increases* with ε growth.

A simple way to estimate the $q$ change vs. ε is to present $\partial\log(\sigma)/\partial\log(q^*)|_\varepsilon$ as $(\mu/\mu_{300})(\sigma_{300}/\sigma)$ [34,43]. Taking $(\mu/\mu_{300})=5$ [32], $(\sigma_{300}/\sigma)=120$, and $d\log(\sigma)/d\varepsilon\approx200$ (from fig. 2), from (5) we obtain $d\log(q^*)/d\varepsilon\approx200/(5*120)=0.33$, or $g=-d\log(q)/d\varepsilon-1=-d\log(q^*)/d\varepsilon=-0.33$. This value of $g$ appears 20 times below the value inferred earlier [15] and is of another sign. This means that under, say, 1% stretch λ/$c$ decreases by 0.3 %. This only drives the CDW away from 4-fold commensurability. With this $q^*(\varepsilon)$ dependence one cannot attribute the anomalies in σ(ε) to a lock-in transition, at least, if 4-fold commensurability is implied.

A similar experiment was performed for $K_{0.3}MoO_3$ – the blue bronze (BB), for which $T_P$=180 K. To the best of our knowledge, BB has not been studied under uniaxial strain before. The techniques applied to the whiskers [3,4,5,6] do not work for the bulk crystals of BB. However, our method appeared applicable to thin microcrystals of BB. A lamella of submicron thickness was put on the epoxy substrate covered with a thin layer of semi-liquid epoxy. After final solidification of the epoxy the crystal appeared to be tightly fixed at the substrate, while the upper surface remained above the epoxy. This allowed deposition of gold contacts on the crystal. Bending the substrate in the proper direction one could both stretch and compress the sample. Here we present the ε dependence of σ at $T$=96 K, somewhat above $T_P$/2, like for TaS$_3$. One can see a hysteresis loop very similar with that for TaS$_3$.

A similar analysis is applicable to BB if we consider the "hole" $q$-vector, i.e. treat the 3/4 – filled electronic bands as quarter-filled bands of holes. Similarly with TaS$_3$, $q$ is slightly above $0.25b^*$ [31] and decreases with temperature decrease [44,45]. Correspondingly, the quasiparticles show n-type conductivity. With these reserves, we conclude that the $q$-vector in BB increases with strain, also departing from the $0.25b^*$ value. Taking $(\mu/\mu_{300})\sim1.5$ [46], $(\sigma_{300}/\sigma)=32$, and $d\log(\sigma)/d\varepsilon\approx20$ (from Fig. 3), from (5) we obtain $d\log(q^*)/d\varepsilon\approx20/(32*1.5)=0.4$, or $g=-d\log(q)/d\varepsilon-1=-d\log(q^*)/d\varepsilon=-0.4$

In view of this result, it becomes unclear, why BB does not show elastic anomalies [23], in contrast with TaS$_3$ and other compounds.

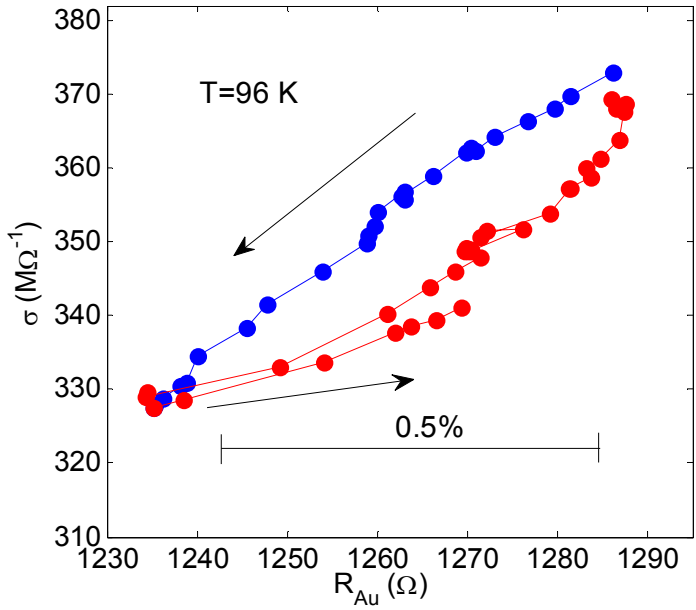

Fig. 3 The $\sigma(\varepsilon)$ loop for BB. The sample length and width are 62 μm and 13 μm respectively.

Before turning to the $\sigma(\varepsilon)$ dependences in nanosized $TaS_3$ samples, we would like to check, if the technique revealing the $q^*$ change based on the "quantization" [35,36] can be applied to $TaS_3$. Though steps for $TaS_3$ nanosamples were observed previously [34,43], discrete equidistant conducting states and regular switching between them with temperature were not clearly demonstrated.

Fig. 4 shows a repeatedly recorded $\sigma(T)$ dependence for a $TaS_3$ nanosample ($s=0.0038$ μm$^2$). One can see that the sample passes discrete conducting states.

Stretching $TaS_3$ beyond the critical value results in a drastic increase of coherence of CDW *sliding* [3,4,5,39]. The coherence of the CDW in the *pinned* state is also found to grow. This can be seen from Fig. 5 where $\sigma(T)$ curves are shown for a sample with dimensions 16 μm×0.3 μm$^2$ at different strain values. The step structure becomes clearly visible at $\varepsilon \geq \varepsilon_c$. This means that the transverse coherence of the CDW increases with strain. Multiply recorded $\sigma(T)$ curve for this sample is also shown in Fig. 4. Note also the lower slope of the $\sigma(T)$ curves between the steps for the stretched sample. Within the semiconductor model [28,34] this means that the quasiparticle conductivity of the stretched sample is "more unipolar". For the case of p-type conductivity this means downward shift of the chemical potential from the gap center, which again corresponds with increase of $q^*$.

From Fig 5 one can also see more pronounced metallic behavior above $T_P$. The transition for $\varepsilon > \varepsilon_c$ becomes more abrupt, and $T_P$ tends to increase in comparison with the value at $\varepsilon < \varepsilon_c$. One can suppose that the $\Delta$ also grows for $\varepsilon > \varepsilon_c$. These features of $\sigma(T)$ indicate reduction of 1D fluctuations with $\varepsilon$ growth and match the formation of ultra coherent CDW (UC CDW) at $\varepsilon > \varepsilon_c$ [39]

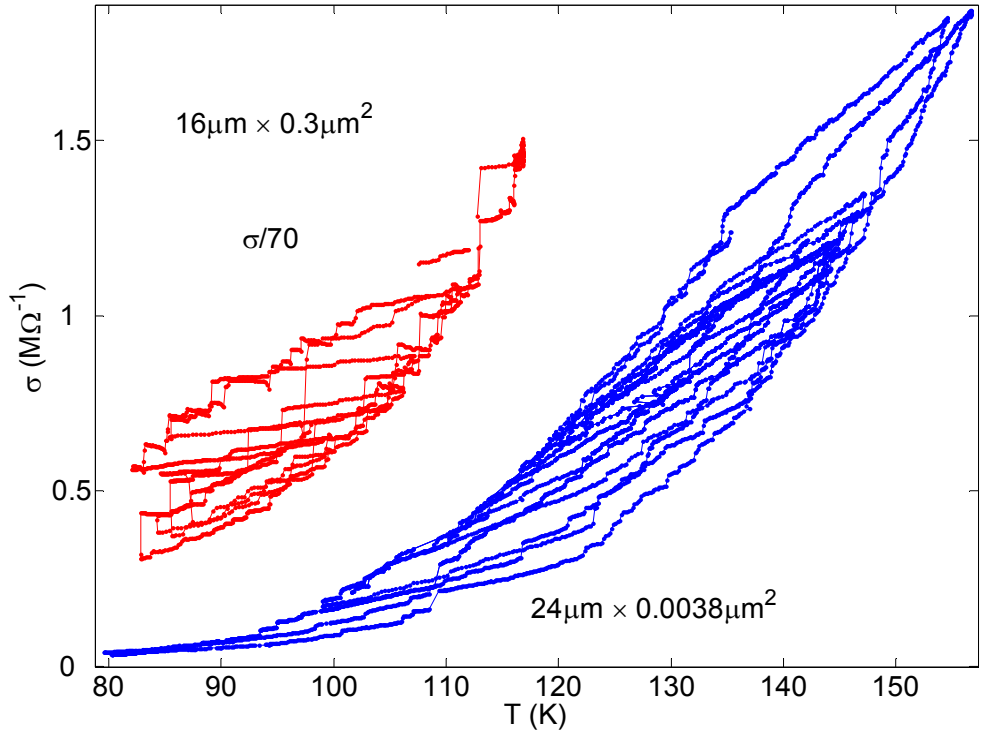

Fig. 4. Repeatedly recorded σ vs. *T* curves for an unstrained (below) and a stretched (above) samples. Note the larger cross-section of the strained sample; its conductivity is divided by 70.

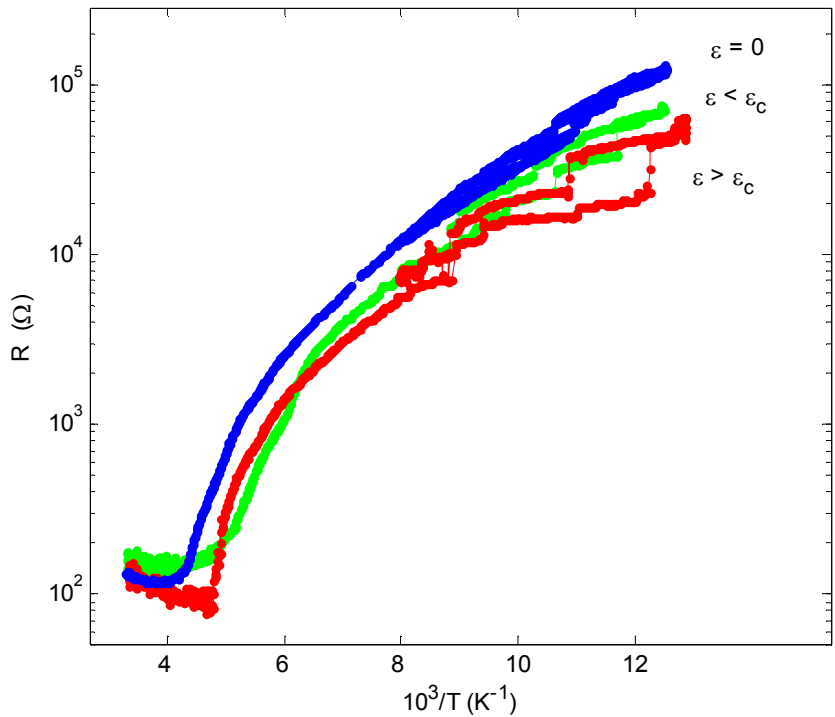

Fig. 5. σ(*T*) curves for the sample with dimensions 16 μm × 0.3 μm² (see also Fig. 4) at ε=0, ε~0.5% (below $\varepsilon_c$) and ε~1% (above $\varepsilon_c$). The sample is prepared with the "lever" method. The resistance of gold film evaporated on top of the sample (280 Ω) is subtracted.

Counting the steps we directly restore the $q^*(T)$ dependence (and its hysteresis) with only one fitting parameter, $q^*(0)$ (Fig. 6). For this calculation we first established the value of the unit step $\delta\sigma \approx 1.2 \cdot 10^{-8}$ Ω corresponding to $2\pi$ phase gain. The majority of the steps were of this kind. Several steps were considered as multiple (up to 4 wavelengths addition/removal), some – as fractional [38]. The data are in agreement with [30]. In particular, extrapolation of the line to $T$=200 K would give $\delta q^*/q^* \approx$ 1-2 %, in accordance with [30]. The slope of the line is about 800 K, which is close to $\Delta$, in agreement with the models [27,28]. From this we conclude that one can determine $q^*$ change in $TaS_3$ nanosamples counting the steps, like for the cases of BB [35] and $NbSe_3$ [36].

Noteworthy is that from the step height, $\delta\sigma$, one can estimate the mobility of the quasiparticles [35,36]. Each phase slip produces or annihilates a given number of electrons or holes, namely, 2 per CDW chain. Therefore, knowing the $\sigma$ change, one can directly determine the mobility. To find $\mu$ one should know the number of conducting chains, $N$, in the sample. It can be found from the Shapiro steps, particularly, from the ratio $I_c/f$:

$$N=(I_c/f)/2e, \qquad (6)$$

where $I_c$ is the CDW current at the Shapiro step. Combing (6) with formula (1) from [35] (see [47]), we obtain:

$$\mu=\delta\sigma L^2/(I_c/f). \qquad (7)$$

This relation appears rather convenient for determination of mobility, because all the factors in (7) can be determined directly from experiment (note that $\sigma$ is total conductivity, not the specific one). For the 24 μm long $TaS_3$, sample from Fig. 4, we found $\mu$ to be 25 cm$^2$/Vs, for the stretched 16 μm long sample from the same picture we find $\mu$ = 33 cm$^2$/Vs, in agreement with the value found from Hall effect studies [32]. This independently confirms that the steps of $\sigma$ are coupled with $q$ change by $2\pi/L$ ($q^*$ change by $c/L$).

Note also that an upward step of conductivity corresponds with the growth of $q^*$, and *vice versa*: only within this assumption the curves in Fig. 6 can be derived from steps counting (Fig. 4). It is known that the steps are directed towards the center of hysteresis loop. Thus, with $\varepsilon$ growth the upward $\sigma$ steps are expected.

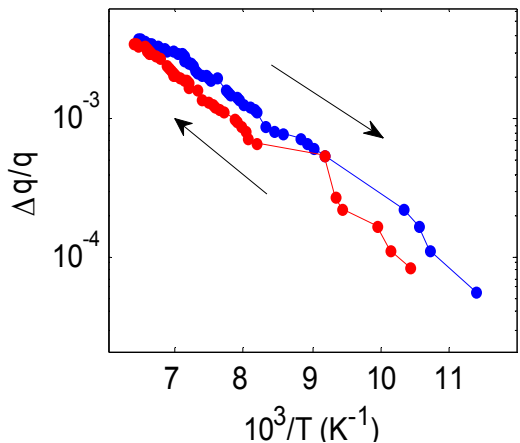

Fig. 6. Normalized $q^*(T)$ dependence obtained counting the steps for the nanosample with dimensions 24μm×0.0038 μm$^2$ (Fig. 4).

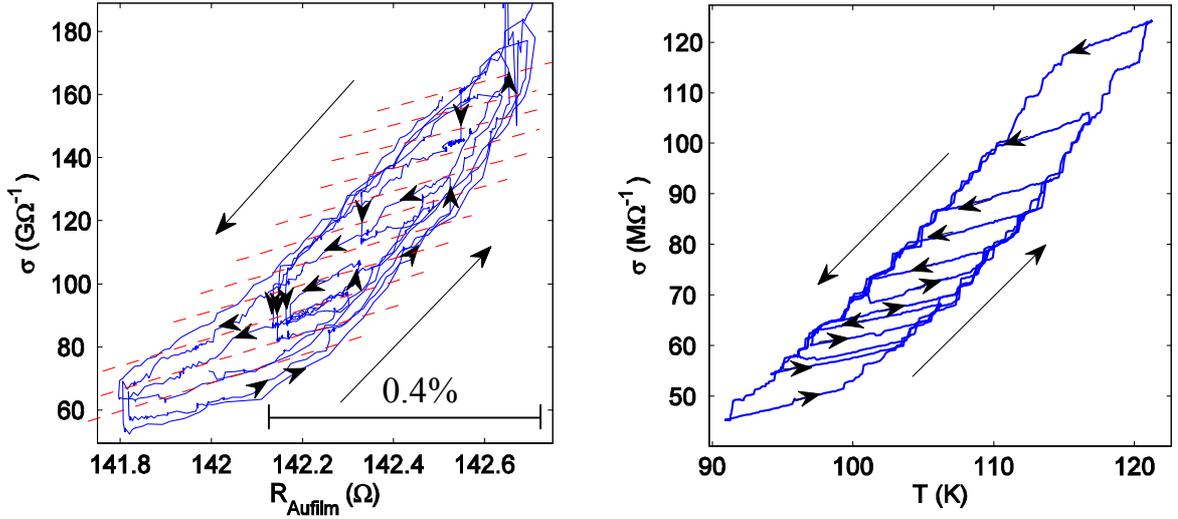

Fig. 7. a) Repeatedly recorded $\sigma(\varepsilon)$ dependences for a TaS$_3$ nanosample. The sample dimensions are 23 μm × (23×10$^{-3}$) μm$^2$. b) Similar dependences vs. *T* for a sample strained above $\varepsilon_c$. The sample dimensions are 19 μm × 0.7 μm$^2$.

Figure 7a shows a hysteresis loop, $\sigma(\varepsilon)$, obtained for a nanosized TaS$_3$ sample. In this experiment an electric motor was used to provide smooth rotation of the micrometer screw and, consequently, very gradual bar bending. The loop looks very similar to the $\sigma(T)$ loops obtained for the nanosized samples (Fig. 4, Fig. 7b). One can distinguish discrete conducting states separated by about 7 GΩ$^{-1}$. Within these states $\sigma(\varepsilon)$ is reversible in $\varepsilon$. The transitions between the states are stepwise. The change of $\varepsilon$ by about 0.25% results in about 10 steps of $\sigma$, i.e. in $\delta q^*/q^*= 10\lambda/L=5.6\times10^{-4}$. From this we find $g= -(\delta q^*/q^*)/\delta\varepsilon = -0.22$. This value of $g$ is close to the found above.

Thus, counting the steps we obtain $q(\varepsilon)$ dependence roughly similar with that for a bulk sample.

## IV. DISCUSSION

The unexpected change of $q$ with $\varepsilon$ has serious consequences. Let us first discuss the $L(T)$ loop for TaS$_3$ [15] in this context. In [15] it was found that in the overcooled state the TaS$_3$ samples are shorter than in the overheated one. Correspondingly, application of voltage to the overcooled sample results in increase of length, and *vice versa* for the overheated state [15]. The hysteresis in $L(T)$ is of unusual kind: thermal expansion goes so, that the length is in front of its equilibrium value, $L_{eq}$. This is not surprising in itself: the length change results from superposition of regular thermal expansion and the effect of CDW strain on top of it. The latter is responsible for hysteresis. Therefore, one cannot tell in advance the appearance of the $L(T)$ loop. However, if we suppose that the sample length change is coupled to the *longitudinal* CDW strain, we will come to a puzzling result: it appears that $L$ change induced by the CDW strain results in further increase of the CDW strain.

Particularly, let us consider the overheated state. With heating $q^*$ increases [30] and, because of hysteresis, is below its equilibrium value, $q^*_{eq}$. At the same time $L>L_{eq}(T)$ [15], and this means, as we know now, that $q^*_{eq}$ is larger than in case of $L=L_{eq}$. The paradoxical result is that while $q^*$ appears below $q^*_{eq}$, the length changes so that $q^*_{eq}$ increases, i.e. the difference $|q^*-q^*_{eq}|$ grows. We come to the apparent contradiction with the very general Le Chatelier's principle: "when a system at equilibrium is subjected to change in concentration, temperature, volume, or pressure, then the system readjusts itself to counteract the effect of the applied change and a new equilibrium is established". In our case $L$ appears to readjust itself so that the CDW longitudinal strain grows.

The most reasonable way to overcome this contradiction is to go beyond the 1D model of CDW-lattice interaction. In [30] the $b^*$ component of the $q$-vector was also found to be strongly temperature dependent, as well as the $c^*$ component. One can suppose that the metastability in the sample dimensions is driven by the CDW-lattice interaction in the direction normal to the chains. Then the length change is to be calculated with the help of Poisson coefficient. It is known that transverse CDW strains dominate over the longitudinal ones (see, e.g. [48]). The enormous voltage-induced torsional strain observed in TaS$_3$ [20] might also indicate that the CDW-lattice interactions, other than the longitudinal uniaxial one, are dominating.

If metastable states in conductivity are dominated by longitudinal CDW strains, while those in length – by transversal CDW strains, one can expect violation of scaling between hysteresis in length and in conductivity. In fact, in [16], where the metastable states were induced by electric field, in the hysteresis loops $L(E)$ and $\sigma(E)$ larger $L$ corresponded to higher $\sigma$, in contrast with [15]. It was also noted [16], that the scaling between $L$ and $\sigma$ was rough; particularly, substantial length changes below $E_t$ were not accompanied by any changes in $\sigma$.

Being responsible for $L$ changes, the transversal components of the CDW strain should be also taken into account in the treatment of the elastic anomalies in the Young modulus. In particular, in the model [29] the coefficient $g$ should be considered as a tensor. Another relevant complication can be connected with the possible coexistence of commensurate and incommensurate CDWs [49].

The obtained for BB $q(\varepsilon)$ dependence indicates that the CDW-lattice interaction is to be reconsidered for this compound as well. Note that torsional strain was observed in this compound, also indicating importance of transversal deformations in the CDW-lattice interaction [18].

Another problem to be solved is the origin of the anomalies at $\varepsilon_c$ [3,4,5,6,39]. The answer may root in the $\sigma(\varepsilon)$ curves (Fig. 2). With increase of $\varepsilon$ from zero up to $\varepsilon_c$ the conductivity grows 2-6 times [3,4,5,6,39]. The hysteresis has allowed to distinguish the changes of $\sigma$ through the $q^*$ dependence on $\varepsilon$ and directly as a function of $\varepsilon$ at fixed $q^*$. The 1st (larger) part of the $\sigma$ change is coupled with $q^*$ increase and corresponding growth of hole concentration. The 2$^{nd}$ (smaller) part can be attributed to the $T_P$ decrease [39,50]. Expansion of quasi one-dimensional conductor along the conducting chains results in a reduction of one-dimensional properties (anisotropy): the distance between lattice cites in the in-chain direction increases, while interchain coupling grows due to the Poisson contraction. The $T_P$ decrease is confirming this. From general consideration, the growth of $q^*$-0.25, i.e., departure of $q$ from the 1D case, also can indicate that the sample becomes "more 3D". Other words, the large growth of quasiparticle conductivity denotes a reduction of quality of nesting of the Fermi surfaces. At $\varepsilon$ close to $\varepsilon_c$, but below it, considerably less number of electrons are gapped by the CDW than at $\varepsilon=0$. Correspondingly, the electronic energy gain due to the lattice distortion also becomes less. However, as one can suppose, at a certain value of $\varepsilon$ a different $q$-vector becomes more effective in gapping electrons. Evidently, the corrugated Fermi surfaces allow a new nesting.

An additional reason for formation of the new CDW can root in the complicated dependence of the electron energy gain on the CDW (lattice distortion) amplitude. Decrease of 1D fluctuations in a stretched sample promotes formation of a higher-amplitude CDW. As it has been noted in [51], for large-amplitude CDWs the nesting is not so crucial: the electrons near $k_F$ would be gapped by the CDW distortion, even if their energy deviates from $E_F$, but the deviation is less than $\Delta$. In this case the electronic energy gain can spread over the entire Brillouin zone [51]. Formally this means that while at low $\Delta$ the energy gain is proportional to $-\Delta^2\ln\Delta$, for high $\Delta$ it becomes proportional to $\Delta$ with a relatively large factor [51]. Presumably, at $\varepsilon=\varepsilon_c$ the minimum of the total energy, i.e. the sum of the positive elastic and the negative electronic energies, can be achieved at two different values of $\Delta$. Above $\varepsilon_c$ the larger $\Delta$ becomes energetically favorable. One can expect that the switching of q-vector at $\varepsilon=\varepsilon_c$ will have features of a 1st-order transition.

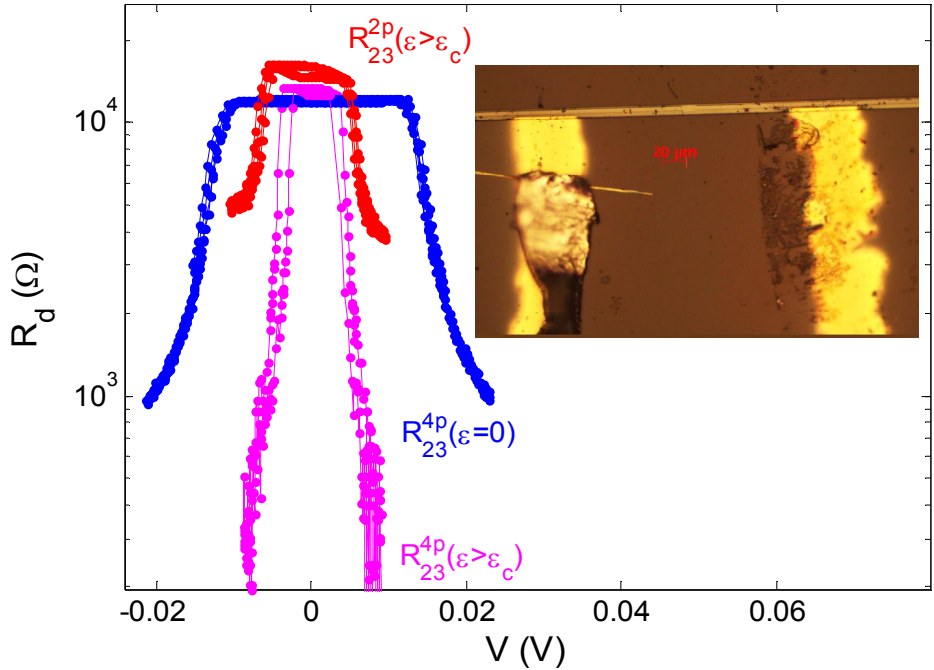

Fig.8. $dV/dI$ vs. $V$ for a TaS$_3$ sample. Results of 2-probe ($\varepsilon>\varepsilon_c$) and 4-probe ($\varepsilon=0$, $\varepsilon>\varepsilon_c$) measurements are shown. $T$=120 K. *Inset*: a microphotograph of the sample fragment. The potential probes ("2" and "3") are shown, their centers are separated by 225 μm. The distance between current probes ("1" and "4", beyond the photo) is 400 μm, $s$=5 μm$^2$.

Formation of the new CDW results in a nearly step-like reduction of σ vs. strain [39], which can indicate both increase of the Peierls gap and partial removal of the electron-hole misbalance. In [14] a feature in the stress-strain dependence, namely, a drop of Young modulus, has been observed at $\varepsilon=\varepsilon_c$. It was explained in terms of a weakly first-order transition of the CDW into a different phase [14]. The authors remarked strongly different properties of the new CDW state forming above $\varepsilon_c$, which could scarcely be explained in terms of a lock-in transition [14]. Evidently, formation of the UC CDW is marked by a gain of electronic energy. This could explain the drop of Young modulus of TaS$_3$ at $\varepsilon_c$ [14].

Though the scenario proposed gives a reasonable explanation of the step-wise changes of the CDW properties at $\varepsilon_c$, it does not give a idea of the extremely high coherence of the new CDW. The features of the UC CDW in linear conduction are seen from Fig 5; the non-linear transport of the UC CDW has been summarized in [39]. The growth of non-linear conduction of the UC CDW at the threshold voltage is even more drastic, than it was reported in [39]: in the 2 probe configuration it is always masked by the contacts resistance. Presumably, the large contacts resistance of the UC CDW is dominated by the phase-slip voltage, which is increased above $\varepsilon_c$. To reduce this contribution to resistance we have measured the differential I-V curves in the 4-probe configuration. For the potential probes the contact film was deposited only at the side (narrow) surface of the sample (inset to Fig. 8), so that shunting of the current by the probes was minimized. One can see

(Fig. 8) that these measurements reveal a drop of the CDW resistance by nearly 2 orders of magnitude in a narrow voltage range. At higher voltages the resistance rapidly approaches the normal state value, but does not exceed it.

The unique features of the UC CDW are still waiting for a reasonable explanation. Note, that the growth of the CDW coherence under stretch can be a universal phenomenon for different CDW conductors: a similar tendency, though without features of a phase transition, was observed for $NbS_3$ (phase II) at room temperature [39].

In conclusion, we report hysteresis in the conductivity-strain dependences for $TaS_3$ and $K_{0.3}MoO_3$ in the CDW state. From the analyses of the loops we found the variation of the longitudinal component of the q-vectors with strain, $\varepsilon$, in these compounds. The "quantization" of $q$ in nanosized $TaS_3$ samples has given another, more direct, way to find $q(\varepsilon)$. In $TaS_3$ $q$ changes so that the CDW departs from $0.25c^*$, and in BB – from $0.25b^*$. In both compounds the electron-hole misbalance grows. This result obliges us to reconsider a number of mechanical effects found for $TaS_3$: metastable length states, drop of the Young modulus on the CDW depinning and strain-induced anomalies in transport properties at $\varepsilon=\varepsilon_c$. Presumably, the latter should be attributed to formation of a new CDW with a larger amplitude and a new $q$-vector providing a nesting the distorted Fermi surfaces. Also, one should find a new explanation of the $K_{0.3}MoO_3$ behavior, which shows no drop of the Young modulus when CDW is depinned. Evidently, to solve these entire problems one should take into account the transverse components in the CDW and lattice deformations.

We are grateful to R.E. Thorne for granting the high-quality $TaS_3$ samples, to A.A. Sinchenko for the help in experiment and to I.G. Gorlova and S.V. Zaitsev-Zotov for fruitful discussions. The support of RFBR (grants 14-02-01240, 14-02-92015, 14-02-01236 and 16-02-01095) and program 'New materials and structures' of RAS are acknowledged. The elaboration of the "bending technique" for bulk samples, allowing both uniaxial expansion and contraction was supported by Russian Scientific Foundation (Grant No14-19-01644)

1. P. Monceau, Advances in Physics **61**, 325 (2012)
2. J.W. Brill in "Handbook of Elastic Properties of Solids, Liquids, and Gases". Vol. II "Elastic Properties of Solids: Theory, Elements and Compounds, Novel Materials, Alloys, and Building Materials" (Ed. M Levy) (San Diego: Academic Press, 2001) Ch. 10, p. 143.
3. V. B. Preobrazhensky, A. N. Taldenkov, and I. Yu. Kal'nova, Pis'ma Zh. Éksp. Teor. Fiz. **40**, 183 (1984) [JETP Lett. **40**, 944 (1984)];
4. V. B. Preobrazhensky, A. N. Taldenkov and S. Yu. Shabanov, Solid State Commun. **54**, 1399 (1985)
5. T. A. Davis, W. Schaffer, M. J. Skove, and E. P. Stillwell, Phys. Rev. B **39**, 10094 (1989)
6. Z. G. Xu and J. W. Brill, Phys. Rev. B **43** 11037 (1991)
7. J. W. Brill and W. Roark, Phys. Rev. Lett. **53**, 846 (1984).
8. L. C. Bourne, M. S. Sherwin, and A. Zettl, Phys. Rev. Lett. **56**, 1952 (1986)
9. Z.G. Xu and J.W. Brill Phys. Rev. B **45**, 3953 (1992)
10. Z. G. Xu and J. W. Brill Phys. Rev. B **43,** 11037 (1991)
11. R.L. Jacobsen, M.B. Weissman, and G. Mozurkewich, Phys. Rev. B **43**, 13198 (1991)
12. X.-D. Xiang and J.W. Brill, Phys. Rev. B **39**, 1290 (1989)
13. X.-D. Xiang and J. W. Brill, Phys. Rev. B **36**, 2969 (1987) (R)
14. K. Das, M. Chung, M. J. Skove, G. X. Tessema, Phys. Rev. B **52**, 7915 (1995)
15. A.V. Golovnya, V.Ya. Pokrovskii, P.M. Shadrin, Phys. Rev. Lett. **88**, 246401 (2002)
16. S. Hoen, B. Burk, A. Zettl, and M. Inui, Phys. Rev. B **46**, 1874 (1992)
17. V. Ya. Pokrovskii, S. G. Zybtsev, and I. G. Gorlova, Phys. Rev. Lett. **98**, 206404 (2007).

18. V.Ya.Pokrovskii, S.G.Zybtsev, V.B. Loginov, V.N. Timofeev, D.V. Kolesov, I.V. Yaminsky and I.G.Gorlova, Physica B **404**, 437 (2009).
19. S. G. Zybtsev, M. V. Nikitin, and V. Ya. Pokrovskii, JETP Letters, **92**, 405 (2010).
20. V. Ya. Pokrovskii, S. G. Zybtsev, M. V. Nikitin, I. G. Gorlova, V. F. Nasretdinova, S. V. Zaitsev-Zotov, Physics-Uspekhi **56**, 29 (2013).
21. G. Mozurkewich, P.M. Chaikin, W.G. Clark, G. Gruner, Solid State Commun. **56**, 421 (1985).
22. X.-D. Xiang, J.W. Brill, Phys. Rev. B **39**, 1290 (1989).
23. L. C. Bourne and A. Zettl, Solid State Commun. **60**, 789 (1986).
24. A. Meerschaut, J. Phys. (Paris) **44**, C3-1615 (1983).
25. Meerschaut & Rouxel, in Crystal Chemistry and Properties of Materials with Quasi-One-Dimensional Structures, 205-279. J. Rouxel (ed.) © 1986 by D. Reidel Publishing Company
26. V.Ya. Pokrovskii, Pisma ZhETF **86**, 290 (2007) [JETP Lett. **86**, 260 (2007)].
27. C. Noguera and J.-P. Pouget, J. Phys. I (Paris) **1**, 1035 (1991).
28. S.N. Artemenko, V.Ya. Pokrovskii, S.V. Zaitsev-Zotov, Zh. Eksp. Teor. Fiz. **110**, 1069–1075 (1996).
29. George Mozurkewich, Phys. Rev. B **42**, 11183 (1990)
30. C. Roucau, J. Phys. (France) C3 **44**, 1725 (1983); Z. Z. Wang, H. Salva, P. Monceau, M. Renard, C. Roucau, R. Ayroles, F. Levy, L. Guemas, and A. Meerschaut, J. Phys. (Paris), Lett. **44**, L315 (1983).
31. In $K_{0.3}MoO_3$ the chains a parellel to the *b* axis
32. Yu. I. Latyshev, Ya. S. Savitskaya, and V. V. Frolov, JETP Lett. **38**, 541 (1983)
33. S. Girault, A. H. Moudden, J. P. Pouget, and J. M. Godard, Phys. Rev. B **38**, 7980 (1988); J. P. Pouget, S. Girault, A. H. Moudden, et al., Phys. Scr. T **25**, 58 (1989).
34. D.V. Borodin, S.V. Zaitsev-Zotov, F.Ya. Nad', Zh. Eksp. Teor. Fiz. **93**, 1394 (1987).
35. S.G. Zybtsev, V.Ya. Pokrovskii, S.V. Zaitsev-Zotov, Nat. Commun. x:x doi: 10.1038/ncomms1087 (2010)
36. S. G. Zybtsev and V. Ya. Pokrovskii, Phys. Rev. B **84**, 085139 (2011)
37. R. Jaramillo, T F. Rosenbaum, E. D. Isaacs, O. G. Shpyrko, P. G. Evans, G. Aeppli, and Z. Cai, Phys. Rev. Lett. **98**, 117206 (2007); R. K. Kummamuru and Y.-A. Soh, Nature **407**, 859 (2008).
38. If the CDW phase at the contacts is not fixed, some phase slips can perturb sample volume from both sides of a contact. Then the resistivity changes both within the studied sample segment and beyond it. In this case fractional steps in $\sigma(T)$ are expected, and counting the number of steps gives the approximate value of $q^*$ change.
39. S.G. Zybtsev and V.Ya. Pokrovskii, Physica B **360**, 34 (2015).
40. E. Slot, M.A. Holst, H.S.J.van der Zant, S.V. Zaitsev-Zotov, Phys. Rev. Lett. **93**, 176602 (2004).
41. S. V. Zaitsev-Zotov, V. Ya. Pokrovskii and P. Monceau, JETP Lett. **73**, 25 (2001)
42. A. N. Taldenkov, 1992-1993, unpublished.
43. D. V. Borodin, S. V. Zaitsev-Zotov, and F. Ya. Nad, Pis'ma Zh. Eksp. Teor. Fiz. **43**, 485 (1986).
44. C. Schlenker, J. Dumas, C. Escribe-Filippini, H. Guyot, In Low-Dimensional Electronic Properties of Molybdenum Bronzes and Oxides. Ed. by C. Schlenker. Kluwer, Dordrecht, 1989, p. 159-257.
45. S. Girault, A. H. Moudden, J.-P. Pouget, J. M. Godard, Phys. Rev. B **38**, 7980 (1988).
46. L. Forró, J. R. Cooper, A. Jánossy, and K. Kamarás, Phys. Rev. B **34,** 9047(R) (1988).
47. $\delta\sigma_s=(2/L)e\mu/s_0$, where $\sigma_s$ is the specific conductivity [35]. With $N=s/s_0$ one obtains Eq. (7).

48. K. Tsutsumi, T. Tamegai, S. Kagoshima, M. Sato, Charge Density Waves in Solids, Lecture Notes in Physics, Volume 217, 1985, pp 17-22.
49. K. Inagaki et al., J. Phys. Soc. Jpn. **77**, 093708 (2008)
50. R.S. Lear, M.J. Skove, E.P. Stillwell, J.W. Brill, Phys. Rev. B. **29**, 5656 (1984)
51. K. Rossnagel, J. Phys.: Condens. Matter **23**, 213001 (2011)